**Risk-mediated transmission dynamics govern epidemic trajectories beyond physical mobility**

Ryosuke Omori[1,#,*], Koichi Ito[2,#], Laith J. Abu-Raddad[3,4,5,6], Yoh Iwasa[7]

[1]Division of Bioinformatics, International Institute for Zoonosis Control, Hokkaido University; Sapporo, 001-0020, Hokkaido, Japan.

[2]Faculty of Science and Engineering, Doshisha University; Kyotanabe, 613-0394, Kyoto, Japan.

[3]Infectious Disease Epidemiology Group, Weill Cornell Medicine-Qatar, Cornell University; Doha, Qatar

[4]Department of Population Health Sciences, Weill Cornell Medicine, Cornell University; New York, USA

[5]Department of Public Health, College of Health Sciences, QU Health, Qatar University; Doha, Qatar

[6]College of Health and Life Sciences, Hamad bin Khalifa University; Doha, Qatar

[7]Department of Biology, Faculty of Science, Kyushu University; Fukuoka, 819-0395, Fukuoka, Japan.

[#]These authors contributed equally to this work.

*To whom correspondence should be addressed:

Ryosuke Omori PhD., Email: omori@czc.hokudai.ac.jp

**Abstract**

Standard epidemiological models rely on physical mobility and policy indicators, which fail when physical movement decouples from actual transmission risk. To address this, we establish a unified theoretical framework governed by risk-mediated transmission dynamics. Rather than treating societal responses as independent phenomenological proxies, we embed the underlying risk-avoidance tendency directly into the transmission mechanism. This is parsimoniously formulated via the Weber-Fechner law as a logarithmically scaled response to disease incidence, alongside behavioral fatigue. Analyzing multi-regional COVID-19 data, our risk-mediated model significantly outperforms traditional frameworks. While mobility metrics merely track movement volume, our approach directly captures unobserved qualitative contact changes, such as mask-wearing. By integrating this intrinsic behavioral principle, our framework provides a robust, mobility-independent baseline for predicting future epidemic trajectories.

**Introduction**

The transmission dynamics of an infectious disease are inseparable from human behavioral adaptation. When faced with an epidemic, populations do not act as passive entities; they continuously assess infection risk and alter their behavior to avoid disease [1, 2]. These autonomous protective changes routinely precede official government interventions and persist long after mandates are lifted [3, 4]. During the COVID-19 pandemic, high-resolution accommodation data demonstrated this preemptive response: individuals began canceling travel plans well before formal states of emergency were declared, reacting directly to public risk information rather than institutional rules [5]. While these risk-avoidance decisions are made at the individual level, they scale collectively to drive macroscopic transmission dynamics.

However, the inherent difficulty of directly quantifying these behavioral processes has led conventional epidemiological models to rely on observable physical proxies. Consequently, recent epidemic modeling has predominantly incorporated societal responses as external forcings, specifically assuming that the stringency of top-down non-pharmaceutical interventions (NPIs) [6, 7] and real-time GPS mobility surveillance [8, 9, 10, 11] linearly dictate transmission rates. While physical mobility metrics provided an invaluable proxy during initial lockdowns, the subsequent decoupling of physical movement from infection trajectories exposed the structural limitations of this approach [9, 12]. This decoupling highlights that human risk-avoidance adapts qualitatively, altering the intensity and nature of social contacts (e.g., mask-wearing) even as absolute physical mobility recovers. Standard mechanistic models, reliant on treating these diverse societal responses as independent external forcings, have therefore struggled to explain the substantial residual variance across different waves and regions [13, 14, 15, 16]. This mismatch suggests that the limitation is not merely a lack of granular data, but a structural omission: by treating societal responses as predetermined inputs, current models fail to capture the dynamic coupling between disease intensity and adaptive human behavior. Therefore, moving beyond a reliance on physical proxies, integrating this root behavioral principle directly into the transmission mechanism is essential to provide a far more accurate and cohesive explanation of epidemic dynamics.

To bridge this gap, we establish a modeling framework that integrates human risk-avoidance and pathogen transmission into a single, unified system governed by risk-mediated transmission dynamics (Fig. 1). Recognizing that populations exhibit self-regulated dynamics [17, 18] shaped by the social amplification of risk [19], we mathematically encapsulate this risk-responsive adaptation using the Weber-Fechner Law [20, 21, 22]. Based on this principle, we posit that the macroscopic societal response scales non-linearly with the logarithm of reported incidence, rather than linear case counts. Importantly, to rigorously evaluate this framework against traditional external-forcing models, our mathematical formulation explicitly separates the effect of formal policy interventions from the autonomous behavioral response. Conceptually, however, we view both factors not as independent phenomena, but as distinct macroscopic manifestations of the same fundamental risk-response, characterized by distinct temporal dynamics: policies act as discrete, delayed societal interventions, whereas behavioral adaptations manifest as continuous, immediate population shifts.

By explicitly isolating this intrinsic behavioral mechanism, our framework rigorously rules out other established epidemiological drivers. Crucially, our analysis specifically targets epidemic phases prior to the widespread emergence of immune-escape variants, ensuring that pathogen evolution does not confound the multi-wave dynamics. We test this framework against COVID-19 waves in seven distinct regions (England, UK; Florida, US; Hong Kong SAR; New York State, US; Qatar; Tokyo, Japan; and Victoria, Australia). By explicitly controlling for environmental seasonality and formal policy interventions, we demonstrate that this risk-responsive behavior is neither a regional idiosyncrasy nor a mere artifact of external forcing. Embedding this risk-mediated feedback into the transmission dynamics alongside a mechanism for behavioral habituation [23, 24], our integrated model explains transmission dynamics significantly better than traditional metrics, even under strict penalties for model complexity (AIC). Furthermore, this framework consistently outperforms models driven by Google Mobility data [25], mechanistically capturing the transmission reductions—such as unobserved masking and distancing—that GPS surveillance misses. Thus, incorporating this risk-mediated mechanism provides a robust, mobility-independent baseline for understanding and predicting pandemic dynamics [26, 27].

## Results

### Superior performance of the risk-mediated transmission model

To identify the fundamental drivers of COVID-19 dynamics, we evaluated the goodness-of-fit of three competing models across seven diverse regions. Comparison of the time-series trajectories (Fig. 2) shows differences in how each model tracks the observed data (black dots). While both the baseline model (red line) and the mobility-based model (blue line) approximate the initial growth phase in specific regions such as England, Florida, and New York (Fig. 2A, B, D), they deviate from the initial dynamics in the other jurisdictions. Furthermore, during subsequent phases, both the baseline and mobility-based models exhibit deviations from the observed curves across multiple regions. In contrast, the risk-mediated transmission model (black line) follows the observed trajectories across the entire period, reproducing the peak timings, peak heights, and decay rates of the waves.

These visual configurations correspond with the statistical goodness-of-fit metrics summarized in Supplementary Table 1. Comparing the two external-forcing frameworks, neither provided a consistently superior fit; the baseline model yielded lower information criteria than the mobility-based model in five regions (England, Hong Kong, New York, Qatar, and Tokyo), whereas the mobility-based model showed a better fit in the remaining two (Florida and Victoria). In contrast, the proposed risk-mediated transmission model, which integrates a logarithmic behavioral response to incidence and habituation, consistently yielded the lowest AIC and BIC values across all seven analyzed regions.

### Regional variations in transmission suppression and mobility decoupling

Decomposing the effective reproduction number ($R_e$) revealed two distinct structural phases in transmission suppression across the analyzed regions (Fig. 3). During the initial exponential growth phase of the pandemic, the reduction in $R_e$ was consistently synchronized with the implementation of stringent policy mandates, as observed in New York and England (Fig. 3A, D). In these jurisdictions, the formal policy intervention term ($M_{\text{policy}}$) accounted for the primary reduction in the transmission rate, reflecting the immediate impact of early interventions.

However, as the epidemic progressed into subsequent phases, the dominant factor driving suppression shifted. In regions such as Tokyo, Florida, Victoria, and Hong Kong (Fig. 3B, C, F, G), the risk-responsive behavioral feedback ($M_{\text{behavior}}$) accounted for the majority of the decrease in $R_e$ during its decay phases, alongside the gradual depletion of susceptible individuals. In these instances, the model indicates that $M_{\text{behavior}}$ was the primary component of the estimated suppression, independent of external factors such as fluctuations in formal policy stringency or seasonal environmental forcing.

Comparing Google Mobility data with the estimated behavioral effect further demonstrated a quantitative decoupling between physical displacement and transmission risk (Fig. 4). During the initial intervention phase, both the mobility index and the model-estimated behavioral reduction aligned, jointly dropping to their lowest points. However, as the pandemic continued, physical mobility indices across all regions gradually returned toward their pre-pandemic baselines. Despite this recovery in movement volume, the model indicates that the transmission reduction effect ($M_{\text{behavior}}$) remained high. This highlights a sustained decoupling where transmission probability remained suppressed even as physical displacement recovered.

**Discrepancies in environmental forcing and the emergence of structural lag**

While the risk-mediated transmission model successfully captured the overarching dynamics in most jurisdictions, a notable discrepancy emerged in Hong Kong and Qatar during their later epidemic phases (Fig. 2C, E). In these two regions, the model predicted an exponential resurgence in cases, driven by a calculated 1.3-fold increase in the effective reproduction number due to a winter decrease in absolute humidity.

However, the observed case data (black dots) during these periods remained stable or continued to decline. This divergence indicates that the macroscopic risk-responsive behavioral term, as formulated in our model, was insufficient to fully offset the fixed seasonal environmental forcing. It suggests the presence of additional, unmodeled suppression mechanisms in these specific regions that maintained low transmission despite favorable environmental conditions for the virus.

**Discussion**

A central conclusion of our study is that human epidemic dynamics are fundamentally governed by intrinsic risk-response rather than mere physical displacement. By conceptualizing this behavioral adaptation as a continuous risk-responsive adaptation to perceived infection risk, we show that it captures the transmission reality significantly better than external-forcing models relying on mobility data. The fact that this risk-mediated framework consistently outperforms mobility-based approaches across diverse regions, even under strict penalization for complexity, confirms that these behavioral mechanisms are not statistical artifacts but fundamental epidemiological forces.

Traditional epidemic models often treat humans as passive physical particles whose movement directly dictates transmission. However, our findings demonstrate that the decoupling of physical mobility and epidemic dynamics can be effectively explained by incorporating risk-responsive behavior. During early-stage lockdowns, physical surveillance data remains highly valuable, representing a phase where physical displacement and contact probability are tightly aligned. However, our results highlight that the decoupling of physical movement from actual risk reflects how populations dynamically adjust the nature of social contacts in response to increasing reported cases as a risk factor. This risk-responsive adaptation allows individuals to maintain a high level of suppression through masking, ventilation, and distancing even as physical displacement returns to pre-pandemic levels. In this phase, explicitly modeling the mechanisms of risk-responsive behavioral adaptation allows us to directly capture this qualitative contact reduction, accounting for protective behaviors overlooked by GPS tracking. Therefore, to effectively forecast epidemiological dynamics, it is crucial to construct a broader behavioral framework that encompasses these risk-mediated dynamics. Without this framework, traditional mobility-based models fundamentally misinterpret periods of mobility recovery as imminent resurgences.

In interpreting these regional differences, the statistical dominance of the behavioral term in jurisdictions like Hong Kong and Victoria requires careful contextualization. Our model fundamentally separates formal policy interventions, measured by Oxford Stringency Index [28] (SI), from the risk-responsive behavioral adaptation driven by reported incidence. In practice, however, these two factors are deeply intertwined. In jurisdictions pursuing aggressive containment or Zero-COVID strategies, proactive government risk communication often triggers widespread public behavioral adaptation well before formal SI metrics reach their maximum [29]. Because the public acts swiftly in response to daily case announcements under such regimens, the model structurally attributes the resulting drop in transmission to the risk-responsive behavior rather than the formal policy index. Therefore, a minimal SI contribution does not imply ineffective governance; rather, it suggests that early policy signaling successfully mobilized voluntary adaptations, effectively preempting the need for the prolonged and blunt mandates observed in other regions.

The interpretation of specific point estimates for parameters such as the behavioral lag ($\tau$) and policy implementation delay ($\rho$) warrants careful consideration. Although these two delays represent fundamentally distinct mechanisms, they often exhibit practical non-identifiability in observational settings. In reality, both policy interventions and risk-responsive behavioral changes are driven by fluctuating case numbers. This shared dependency causes simultaneous fluctuations in the data, creating a compensatory ridge in the parameter space (Supplementary Fig. 1). As a result, disentangling these two delays is highly challenging unless sufficient longitudinal data spanning multiple distinct waves is available. Rather than indicating statistical redundancy, the simultaneous inclusion of both delays is a mechanistic prerequisite to capture the dual pathways of transmission suppression. Forcing a mathematically parsimonious model by excluding the risk-responsive behavior ($\tau$) critically distorts the system; it compels the optimization algorithm to compress the policy delay ($\rho \approx 0$) to artificially absorb the preemptive behavioral changes that were already bending the epidemic curve (Supplementary Fig. 2). Therefore, the inclusion of the logarithmic risk-mediated feedback provides the necessary structural flexibility to explain macroscopic multi-wave dynamics without forcing policy variables to compensate for unobserved behavioral adaptations.

The late-phase discrepancies observed in Hong Kong and Qatar can be interpreted as the effect of highly targeted interventions uncaptured by broad, society-wide policy-indices. In these regions, a winter decrease in absolute humidity led the model to predict an environmental resurgence that did not materialize. Because our framework estimates transmission based solely on these population-wide restrictions [28] and risk-responsive behavioral feedback, this divergence indicates the presence of additional suppression mechanisms. Specifically, exhaustive contact tracing and proactive quarantine [30, 31] likely decoupled actual transmission from these universal drivers. Consistent with evidence that climate alone does not determine epidemic dynamics [32] , this discrepancy provides evidence that targeted public health strategies can act as a decisive counter-force to seasonal pressures.

A critical methodological advantage of our framework is its autonomy from empirical behavioral data, such as Google Mobility [25] or survey-based metrics. Most existing mathematical models are input-dependent, requiring external datasets to characterize the population's response [33]. In contrast, our approach models the internal mechanism of the population-level response itself, closing the risk-mediated feedback loop between incidence reports and collective behavioral adjustment [18]. This independence provides a significant strategic advantage for proactive policy assessment and counterfactual simulation. Because the model operates as an autonomous system, it allows public health authorities to simulate hypothetical scenarios before an intervention is implemented. By characterizing the risk-responsive behavioral dynamics, authorities can estimate the net impact of a proposed policy by anticipating how it will interact with the population's existing risk-mediated feedback loop. This enables the optimization of intervention timing and intensity without the delays, biases, or privacy concerns associated with real-time GPS tracking.

From a global health perspective, the data-light nature of our model offers broader applicability for epidemic forecasting. While high-income countries can afford high-resolution mobility tracking, many low-and-middle-income countries lack the digital infrastructure or the privacy frameworks to implement such surveillance [34]. By providing a robust model that requires only case incidence data, we offer a tool that is applicable regardless of a nation's technological standing. Moreover, this approach circumvents the ethical dilemmas associated with mass physical surveillance, providing a privacy-by-design alternative that respects individual autonomy [34]. By prioritizing transparent risk communication over blunt physical mandates, public health strategies can foster more sustainable transmission suppression. This approach directly leverages the intrinsic, risk-mediated feedback between public information and behavioral adaptation.

Our framework presents certain limitations. First, the model assumes a constant reporting rate and antigenic stability. Although we truncated the study period to minimize the confounding effects of viral evolution, temporal variations in surveillance capacity or minor variant emergence could still influence the estimated feedback strength. Second, while the risk-mediated feedback term explicitly captures the macroscopic cognitive-behavioral response to incidence, it cannot disaggregate the relative contributions of specific protective actions, such as mask-wearing, ventilation, or physical distancing. Integrating granular, intervention-specific datasets remains a key challenge for resolving these micro-level behavioral dynamics within risk-mediated epidemiological models.

In conclusion, our study establishes that human epidemic dynamics are fundamentally shaped by risk-responsive behavioral adaptation, extending beyond physical movement patterns. By demonstrating the consistent superiority of a risk-mediated transmission model over mobility-based approaches, we provide a mechanistic foundation for understanding the decoupling of movement and infection risk. The autonomy of our model from empirical mobility data offers a robust and ethically sound framework for proactive policy assessment, allowing for the simulation of intervention outcomes based on risk-responsive behavioral mechanisms. Therefore, incorporating risk-mediated feedback as a primary epidemiological driver provides a critical baseline for pandemic management. Rather than relying exclusively on top-down physical restrictions, future public health strategies can leverage and anticipate the intrinsic, risk-responsive behavioral dynamics of populations.

## Materials and Methods

### Study Design and Data Sources

We analyzed the daily confirmed COVID-19 cases in seven distinct regions: England (UK), Florida (US), Hong Kong SAR, New York State (US), Qatar, Tokyo (Japan) and Victoria (Australia). These regions were selected based on the availability of continuous daily incidence data and corresponding policy stringency metrics at a consistent spatial resolution. The study period was restricted from the onset of the pandemic in early 2020 (2nd March for England and New York, 9th March for other regions) to 1st November 2020 for England and Victoria, 27th December 2020 for other regions. This temporal truncation was strictly selected to ensure antigenic stability. While the D614G substitution emerged during this period, enhancing viral transmissibility, established literature confirms that it did not significantly alter antigenicity or immune escape properties compared to the ancestral strain[35, 36]. By concluding the analysis prior to the global dominance of the Alpha variant (B.1.1.7), we isolated the effects of behavioral changes from complex viral evolution.

**Mathematical Modeling Framework**

To reconstruct the transmission dynamics accounting for the delay between infection and infectiousness, we employed a Susceptible-Exposed-Infected-Recovered (SEIR) model. The population is divided into four compartments: Susceptible ($S$), Exposed ($E$), Infectious ($I$), and Recovered ($R$). The dynamics are governed by the following system of differential equations: $dS/dt = -\lambda S$, $dE/dt = \lambda S - \sigma E$, $dI/dt = \sigma E - \gamma I$, and $dR/dt = \gamma I$, where $\lambda = \beta_t I/N$ is the force of infection. In this model, the effective reproduction number $R_e(t)$ can be written as $\beta_t S/(\gamma N)$.

We fixed the transition rates based on established epidemiological characteristics of the ancestral SARS-CoV-2 strain. We assumed a mean latent period ($1/\sigma$) of 3.7 days [37, 38] and a mean infectious period ($1/\gamma$) of 3.5 days [37, 38]. Crucially, this parameterization sets the latent period to be shorter than the mean clinical incubation period (~5.2 days). This gap accounts for pre-symptomatic transmission, reflecting viral shedding dynamics where a substantial proportion of transmission occurs prior to symptom onset, as reported by He et al. [39]. This structural distinction allows the model to capture the rapid dissemination of the virus driven by undocumented infectious individuals before clinical detection.

The time-varying transmission rate, $\beta_t$, is modeled as the product of a baseline rate ($\beta_0$) and three modifying factors representing climate, policy, and risk-responsive behavior:

$$\beta_t = \beta_0 \times M_{climate}(t) \times M_{policy}(t) \times M_{behavior}(t).$$

**Modeling Seasonal Forcing ($M_{climate}$)**

To account for the environmental dependence of SARS-CoV-2 transmission, we modeled the seasonal forcing term, $M_{climate}(t)$ as:

$$M_{climate}(t) = \exp\big(-\alpha \cdot \mathrm{AH}(t)\big)$$

where $\mathrm{AH}(t)$ denotes the daily absolute humidity ($g/m^3$) [32]. The sensitivity parameter $\alpha = 0.02$ was determined to yield a seasonal transmission reduction consistent with empirical estimates for endemic Betacoronaviruses in temperate regions [40, 41]. The daily absolute humidity data were obtained from the Open-Meteo API [42].

**Policy Intervention ($M_{policy}$)**

For policy intervention ($M_{policy}$), we utilized the Oxford Stringency Index (SI) [28] incorporating a "Pandemic Fatigue" mechanism. We assumed that the efficacy of policy decays with the cumulative duration of restrictions, modeled as:

$$M_{policy}(t) = \exp\left(-\frac{1}{h_a + h_b \sum_{k=0}^{t} SI_k} \cdot SI_{t-\rho}\right)$$

where $h_a$ denotes the baseline efficacy of the intervention, and $h_b$ represents the fatigue rate, creating a tolerance effect reflecting behavioral habituation. Additionally, $SI_{t-\rho}$ represents the Stringency Index evaluated with a time lag of $\rho$ days, which accounts for the inherent delay between the implementation of a policy and its subsequent manifestation in population behavior and transmission dynamics (Fig. 1).

**Risk-responsive Behavior ($M_{behavior}$)**

For the risk-responsive behavior ($M_{behavior}$), we utilized a feedback mechanism grounded in our previous empirical findings. Omori et al. [21] identified that voluntary movement avoidance in Japan scales with the logarithm of reported cases. By integrating this empirical scaling (the Weber-Fechner Law [22]) with a logistic response function, we derived the behavioral reduction factor. The full derivation of this functional form is detailed in the SI Appendix. The resulting equation for the behavioral response is:

$$M_{behavior}(t) = 1 - p\frac{\left(J(t-\tau)\right)^n}{\left(K(t-\tau)\right)^n + \left(J(t-\tau)\right)^n}$$

where $p$ is the maximum reduction rate, $\tau$ is the collective adaptation lag, $K(t-\tau)$ is the habituation threshold representing accumulated fatigue, and $n$ is the coefficient governing the sharpness of the behavioral switch. Crucially, $J(t-\tau)$ represents the daily newly reported cases observed by the public; thus, the lag $\tau$ captures the time required for population-level behavioral adjustments in response to reported information, distinct from the latent epidemiological reality (Fig. 1).

Within this framework, to capture behavioral fatigue (habituation) while maintaining temporal and causal consistency, we modeled the threshold $K$ not as a constant, but as a dynamic variable evaluated at the same lagged time point $(t-\tau)$ as the stimulus. We defined $K(t-\tau)$ as a linear function of the cumulative reported cases up to that point:

$$K(t-\tau) = k_a \sum_{s=0}^{t-\tau} J(s) + k_b.$$

This formulation posits that as the public demonstrates behavioral habituation to the accumulation of cases, the threshold for triggering avoidance ($K$) rises, requiring a higher daily report count to elicit the same level of suppression.

**Model Formulations**

To systematically isolate the explanatory power of risk-responsive behavior, we compared three competing model structures. First, we defined a baseline model serving as a null hypothesis where populations lack autonomous behavioral adaptation, driven exclusively by externally-forced environmental and policy factors:

$$\beta_t = \beta_0 \times M_{climate}(t) \times M_{policy}(t).$$

Second, to validate our framework against standard surveillance-based approaches, we constructed a competing model driven by physical mobility data (mobility-based model). Following the established methodology by Nouvellet et al. [9], we modeled transmission reduction as an exponential function of the Google Community Mobility Index [25] ($G_t$):

$$\beta_t = \beta_0 \times M_{climate}(t) \times \exp(-g \cdot \Delta G_t)$$

where $\Delta G_t$ represents the relative reduction in mobility. Finally, our proposed risk-mediated transmission model replaces the empirical mobility term with the risk-responsive behavior mechanism:

$$\beta_t = \beta_0 \times M_{climate}(t) \times M_{policy}(t) \times M_{behavior}(t).$$

**Observation Process, Reporting Delays, and Parameter Estimation.**

We estimated the model parameters (i.e., $\beta_0$, $\rho$, $\tau$, $p$, $n$, $k_a$, $k_b$, $h_a$, $h_b$, and $g$) using Maximum Likelihood Estimation. To account for the time lag between the moment of infection and the eventual reporting, we explicitly modeled the observation process. We assumed that the daily number of newly reported cases follows a Poisson distribution. The expected number of reported cases for each day was calculated by convolving the daily incidence of new infections (simulated by the SEIR model) with the reporting delay distribution, assuming a constant reporting rate. The initial latent infected population ($I_0$) was co-estimated solely as a nuisance parameter to align the simulated onset with early surveillance data; thus, $I_0$ consumes a degree of freedom in our AIC and BIC calculations but is excluded from Supplementary Table 1. The negative log-likelihood was subsequently minimized using the Nelder-Mead simplex algorithm to fit with the number of reported COVID-19 cases[43, 44, 45]. To mitigate the risk of converging to local optima, the optimization was initialized from multiple random starting points within biologically plausible parameter spaces. The confidence intervals (CIs) of the model parameters were computed using the bootstrap method.

To explicitly model the time lag between infection and case reporting, we defined the total reporting delay probability distribution, $f_{delay}(t)$, as the convolution of the incubation period and the region-specific onset-to-reporting delay (Fig. 1). For the incubation period, we consistently applied a global Log-Normal distribution ($\mu = 1.621, \sigma = 0.418$, corresponding to a median of 5.1 days[46]). Regarding the onset-to-reporting delay, we parameterized region-specific distributions based on local surveillance data. For Tokyo, we used a Log normal distribution with parameters $\mu = 1.72$ and $\sigma = 0.61$, truncated at a maximum of 30 days to account for long-tail delays [47]. Similarly, delays in England, New York and Victoria were modeled using Log-normal distributions with parameters $\mu = 0.842, \sigma = 1.435$ [48], $\mu = 1.95, \sigma = 0.55$[49, 50] and $\mu = 1.39, \sigma = 0.60$[51], respectively. For Hong Kong, we parameterized a Log-normal distribution ($\mu = 1.68, \sigma = 0.63$) to approximate the delay distributions reported by Cowling et al., which ranged from a median of 3 to 9 days depending on the epidemic phase[30]. For Qatar, the onset-to-reporting delay was parameterized by a log-normal distribution with a median of 3 days (IQR: 1–5). This setting is based on high-intensity surveillance data from similar rapid-response environments[52] and aligns with Qatar's reported operational capacity for contact tracing[31]. For Florida, we utilized an estimated delay distribution from infection to reporting derived from 2020 Florida line-list data, which is available in the R package "incidental"[53].

**Model Comparison.**

Model selection across these structures was performed using the Akaike Information Criterion (AIC) and Bayesian Information Criterion (BIC). In calculating the penalty terms, the degrees of freedom strictly incorporated all explicitly estimated structural parameters alongside the initial state nuisance parameter ($I_0$), and the sample size for the BIC was defined as the total number of observation days in the analyzed period for each region. Models yielding lower information criteria were selected as the best-fitting models.

**Derivation of the risk-responsive behavior function.**

To formulate the risk-responsive behavior, we first define the stimulus $x$ induced by the reported number of infected individuals. According to the Weber-Fechner Law, the psychological response to a stimulus scale logarithmically with the ratio of the stimulus to a baseline threshold. Therefore, we define the stimulus $x$ at time $t$ with a time lag $\tau$ as:

$$x = \log\left(\frac{J(t-\tau)}{K(t-\tau)}\right) \quad (1)$$

where $J(t-\tau)$ is the number of newly reported cases observed by the public $\tau$ days ago, and $K(t-\tau)$ is the cumulative burden of the epidemic up to that point, representing the threshold of habituation. The threshold $K$ is modeled as a linear function of the cumulative reported cases:

$$K(t-\tau) = k_a \sum_{s=0}^{t-\tau} J(s) + k_b \quad (2)$$

To ensure the behavioral response is bounded between 0 and 1, we employ a logistic (sigmoid) function:

$$f(x) = \frac{1}{1+\exp(-nx)} \quad (3)$$

where $n$ determines the steepness of the response to the stimulus. The overall behavioral multiplier, $M_{behavior}$, is assumed to be reduced from the baseline value of 1 by a maximum fraction $p$ due to this response. Thus, we have:

$$M_{behavior}(t) = 1 - p \cdot f(x) \quad (4)$$

Substituting equation (1) into (3), and subsequently into (4), yields:

$$M_{behavior}(t) = 1 - p\frac{1}{1+\exp\left[-n\left(\log J(t-\tau) - \log K(t-\tau)\right)\right]} \quad (5)$$

Through simplification, we obtain the final form of the risk-responsive behavior function used in the section "Risk-responsive behavior ($M_{behavior}$)" of material and methods:

$$M_{behavior}(t) = 1 - p\frac{\left(J(t-\tau)\right)^n}{\left(K(t-\tau)\right)^n + \left(J(t-\tau)\right)^n} \quad (6)$$

**Data availability**

All data used in the analyses are available at the references cited in the Methods: the Johns Hopkins University Center for Systems Science and Engineering (JHU CSSE) COVID-19 Data Repository [43], the UK Health Security Agency (UKHSA) data dashboard [44], the Ministry of Health, Labour and Welfare of the Japanese government [45], the Google COVID-19 Community Mobility Reports [25], the Oxford COVID-19 Government Response Tracker (OxCGRT) [28], the Open-Meteo API [54], the R package 'incidental' [53].

**Code availability**

Code for this manuscript is publicly available at https://github.com/hmito/risk-mediated_SIR

**Acknowledgments**

Research reported in this publication was supported by the Japan Society for the promotion of Science, and the Qatar Research Development and Innovation Council. The content is solely the responsibility of the authors and does not necessarily represent the official views of the Japan Society for the promotion of Science and Qatar Research Development and Innovation Council. The funder of the study had no role in study design, data collection, data analysis, data interpretation, or writing of the article. Statements made herein are solely the responsibility of the authors.

**Author Contributions**

R.O. and K.I. designed research; R.O. and K.I. performed research; R.O. and K.I. analyzed data; L.J.A. and Y.I. supervised research; R.O. and L.J.A. acquired funding; and R.O., K.I., L.J.A., and Y.I. wrote the paper.

**Competing Interests**

The authors declare no competing interests.

## Figures

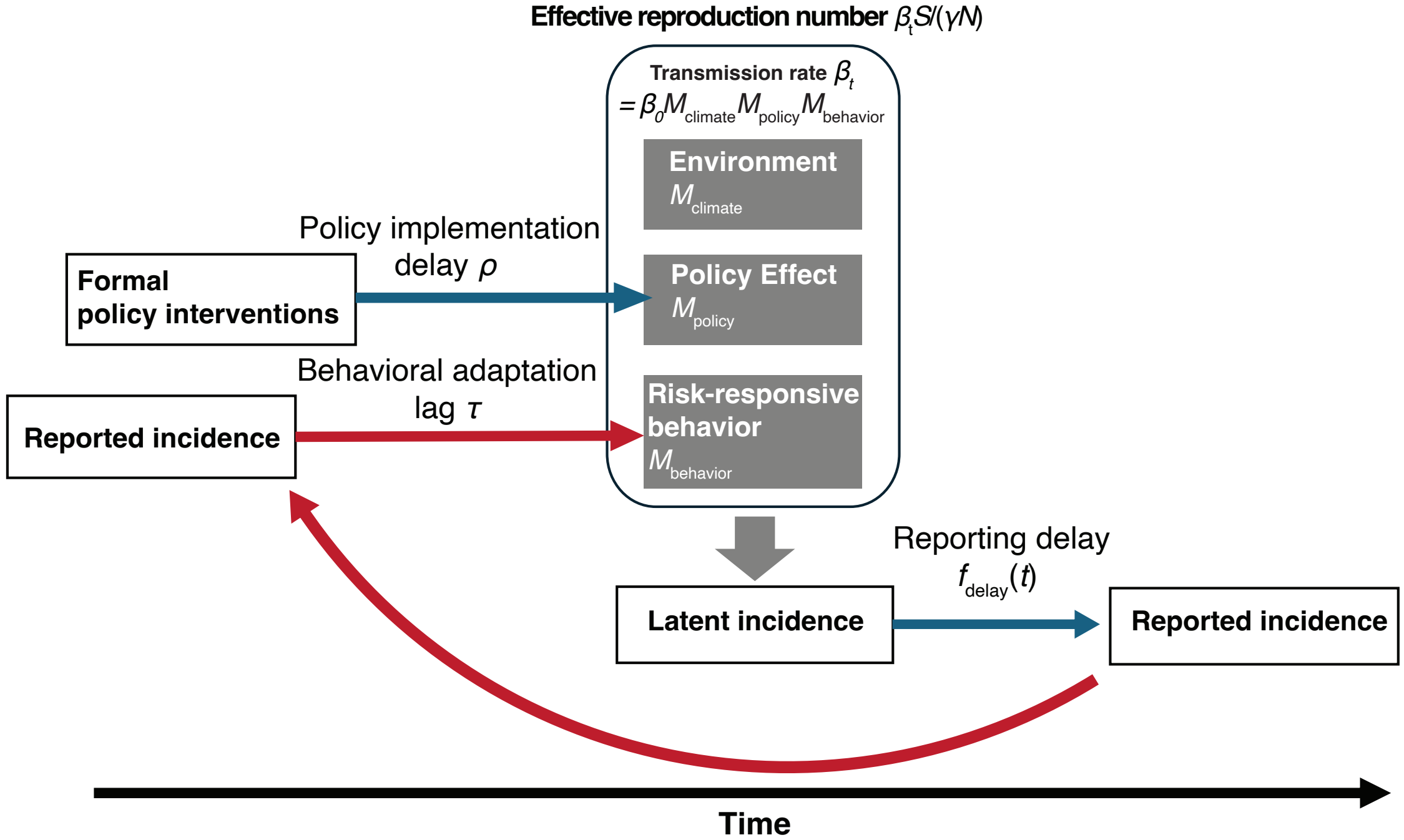

**Figure 1. Schematic representation of the multi-layered delay structure and risk-response mechanisms in the epidemiological dynamics.** The diagram explicitly visualizes how three distinct temporal delays mediate the closed-loop system of viral transmission and societal response. First, the observation filter $f_{delay}(t)$ captures the administrative and biological delay, the convolution of the incubation period and reporting delays, translating actual latent incidence into publicly reported cases. Subsequently, these observed cases drive two parallel feedback pathways, each governed by its own inherent lag. The risk-responsive behavioral adaptation is constrained by a cognitive lag $\tau$, representing the time required for society to perceive the reported risk and adjust collective behavior. Simultaneously, formal policy interventions are constrained by an implementation lag $\rho$. By explicitly mapping these sequential delays along the timeline, the model demonstrates how the temporal decoupling between physical transmission and risk-responsive behavioral adaptation generates complex epidemic waves.

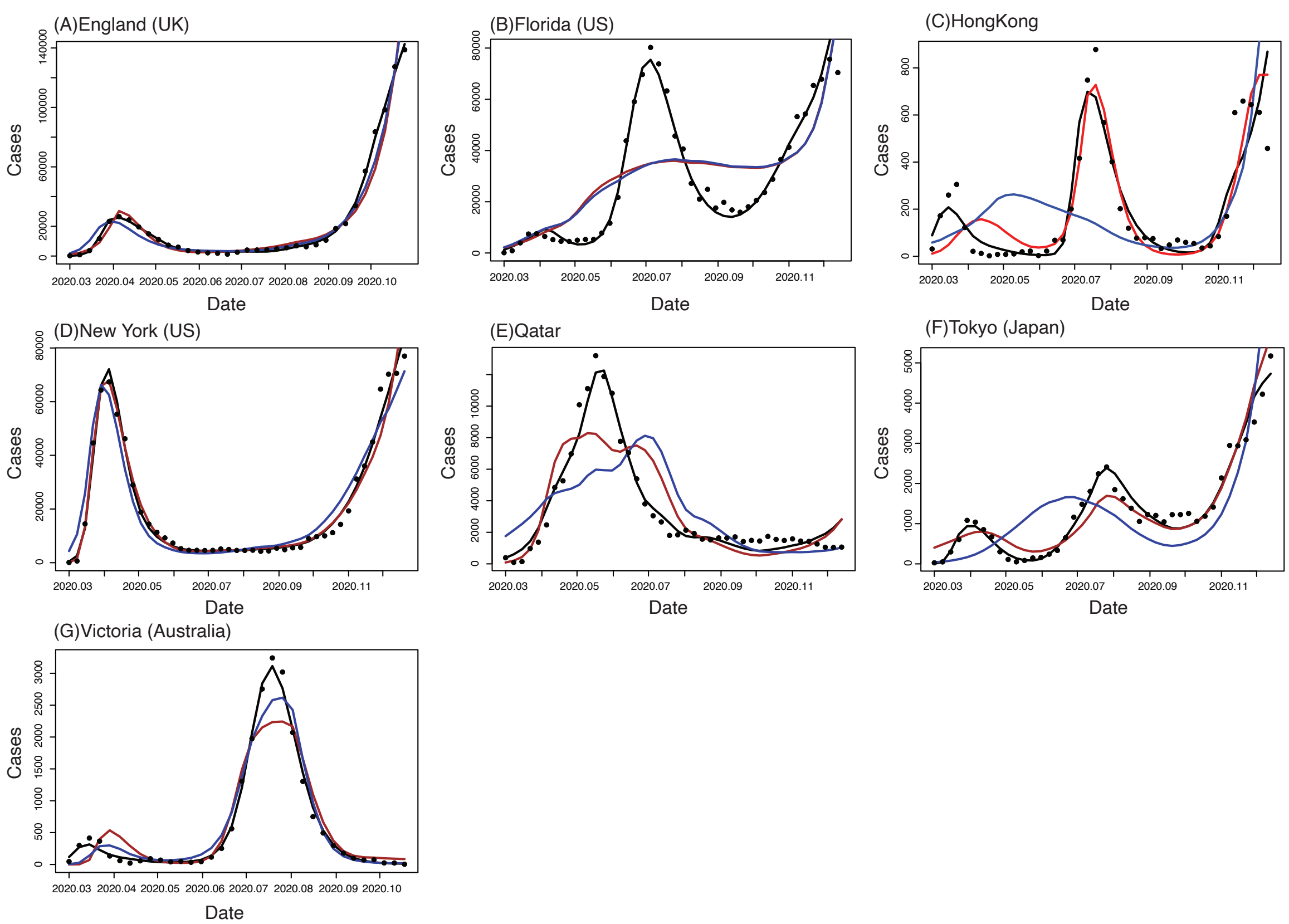

**Figure 2. Model fit of the risk-mediated transmission model across seven regions.** Black dots represent observed daily reported cases. The solid black line indicates the incidence estimated by the risk-mediated transmission model, which integrates environmental forcing, formal policy interventions, and risk-mediated feedback. The solid red line represents a baseline model excluding the risk-responsive behavioral term. The solid blue line denotes the mobility-based model, in which the transmission rate is driven by environmental forcing and empirical Google Mobility data, rather than formal policy interventions and risk-responsive behavioral adaptation.

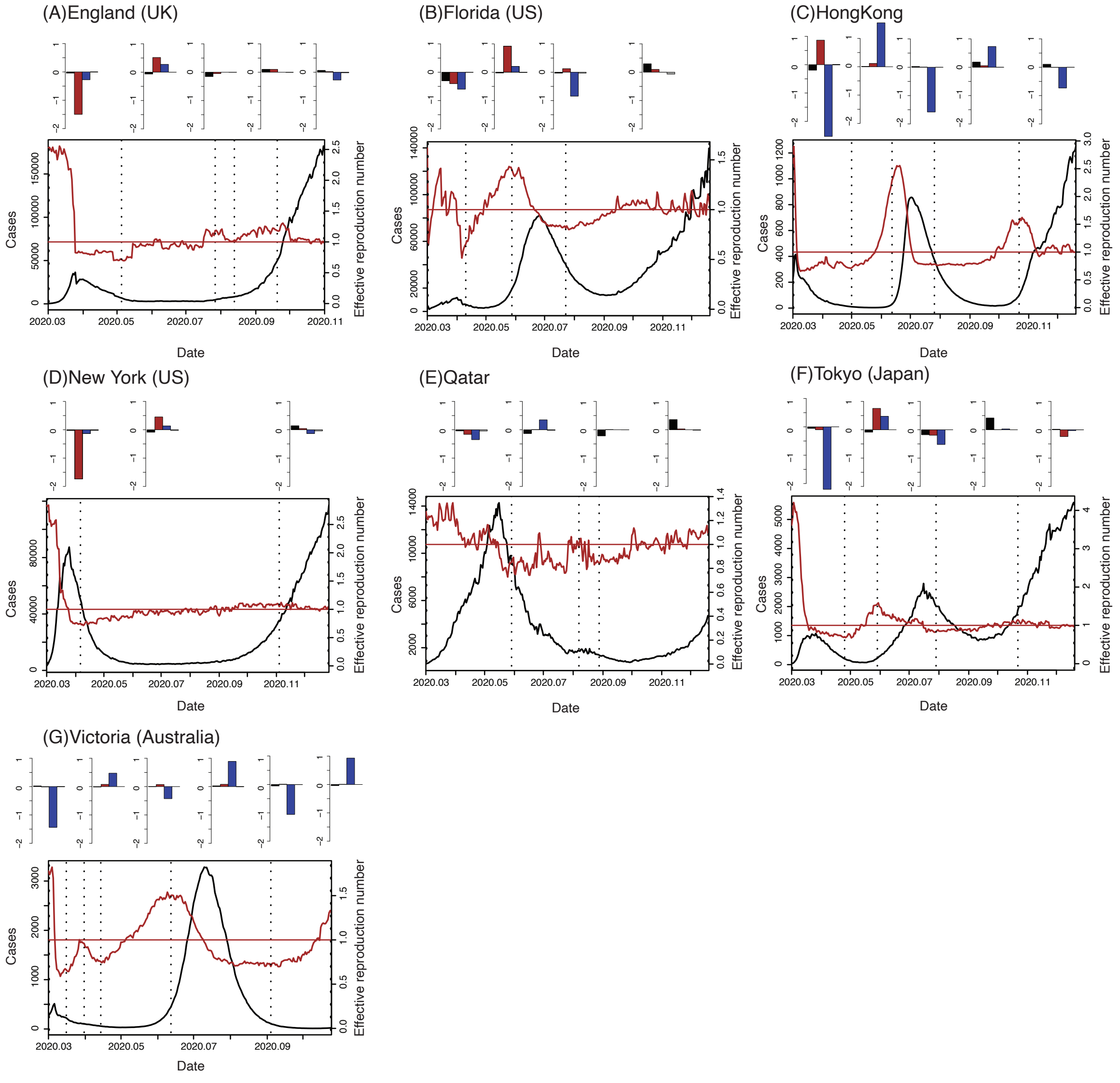

**Figure 3. Phase-dependent transmission dynamics and factor decomposition across seven regions.** Estimated transmission dynamics and contribution of driving factors for England, UK (A), Florida, US (B), Hong Kong (C), New York, US (D), Qatar (E), Tokyo, Japan (F), and Victoria, Australia (G). In the time-series panels, the black line represents the estimated effective reproduction number (right vertical axis) and the red line represents the predicted incidence (left vertical axis). The associated bar charts illustrate the log-transformed contribution rate of four factors to the instantaneous changes in the effective reproduction number: environmental forcing (black), formal policy interventions (red), risk-responsive behavior (blue),

and depletion of susceptibles (gray). Each bar represents the mean contribution of these factors during the respective increase or decrease phases of the effective reproduction number.

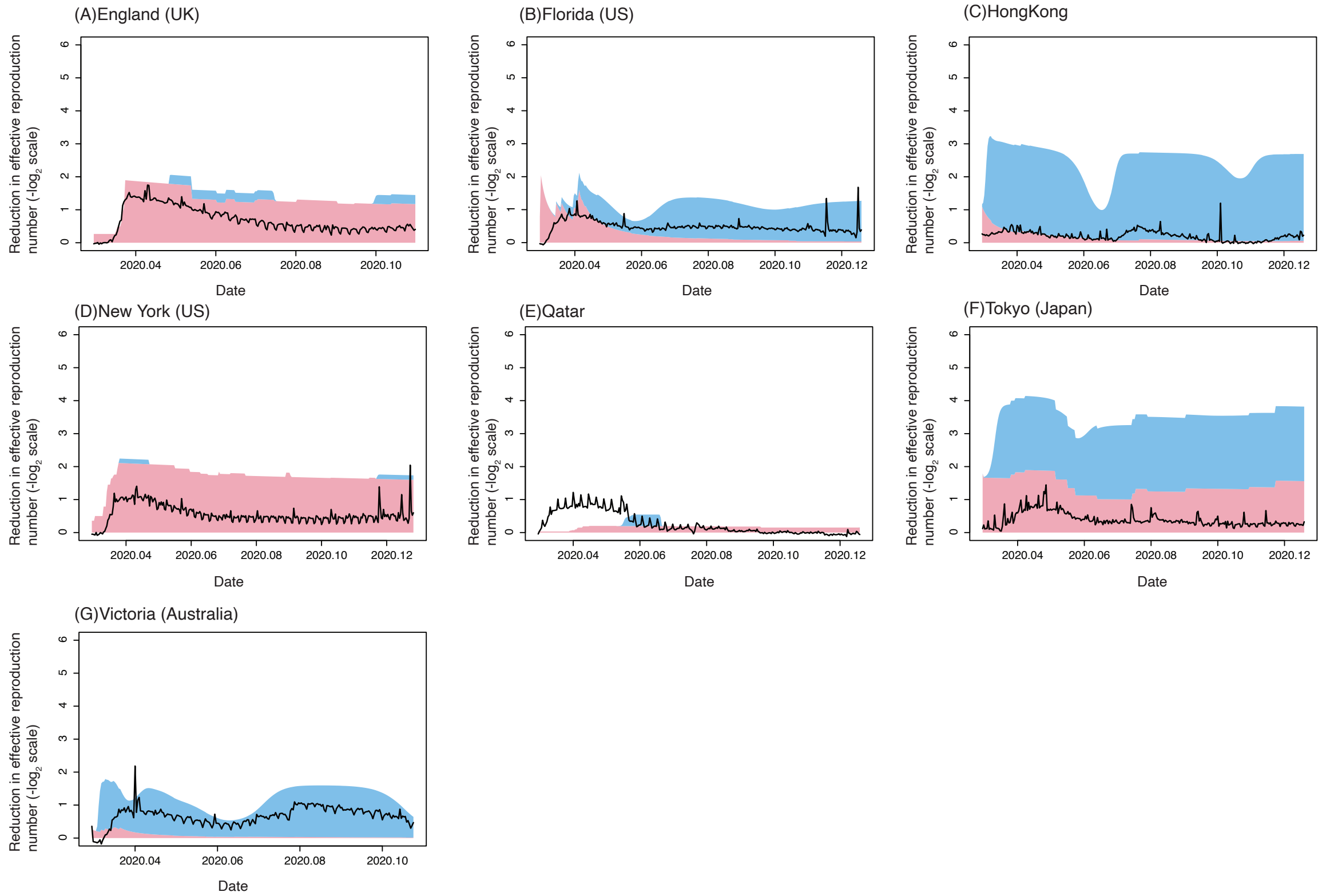

**Figure 4. Decoupling of qualitative behavioral adaptation and physical mobility.** The stacked areas represent the cumulative reduction in the transmission rate attributed to formal policy interventions (red) and risk-responsive behavior (blue). To facilitate direct comparison with mobility data, all reduction factors ($y$) are transformed to a $-\log_2(y)$ scale. On this scale, a value of 1.0 represents a 50% reduction (halving) of the transmission rate, and 2.0 represents a 75% reduction (quartering). The solid black line indicates the reduction in Google Mobility indices relative to the pre-pandemic baseline. The divergence between the mobility line and the total stacked bars highlights the significant role of risk-responsive behavioral adaptations in maintaining transmission suppression even as physical movement returned toward baseline levels.

Supplementary Information for

# Risk-mediated transmission dynamics govern epidemic trajectories beyond physical mobility

Ryosuke Omori, Koichi Ito, Laith J. Abu-Raddad, Yoh Iwasa

Ryosuke Omori

Email: omori@czc.hokudai.ac.jp

## Supplementary Figures

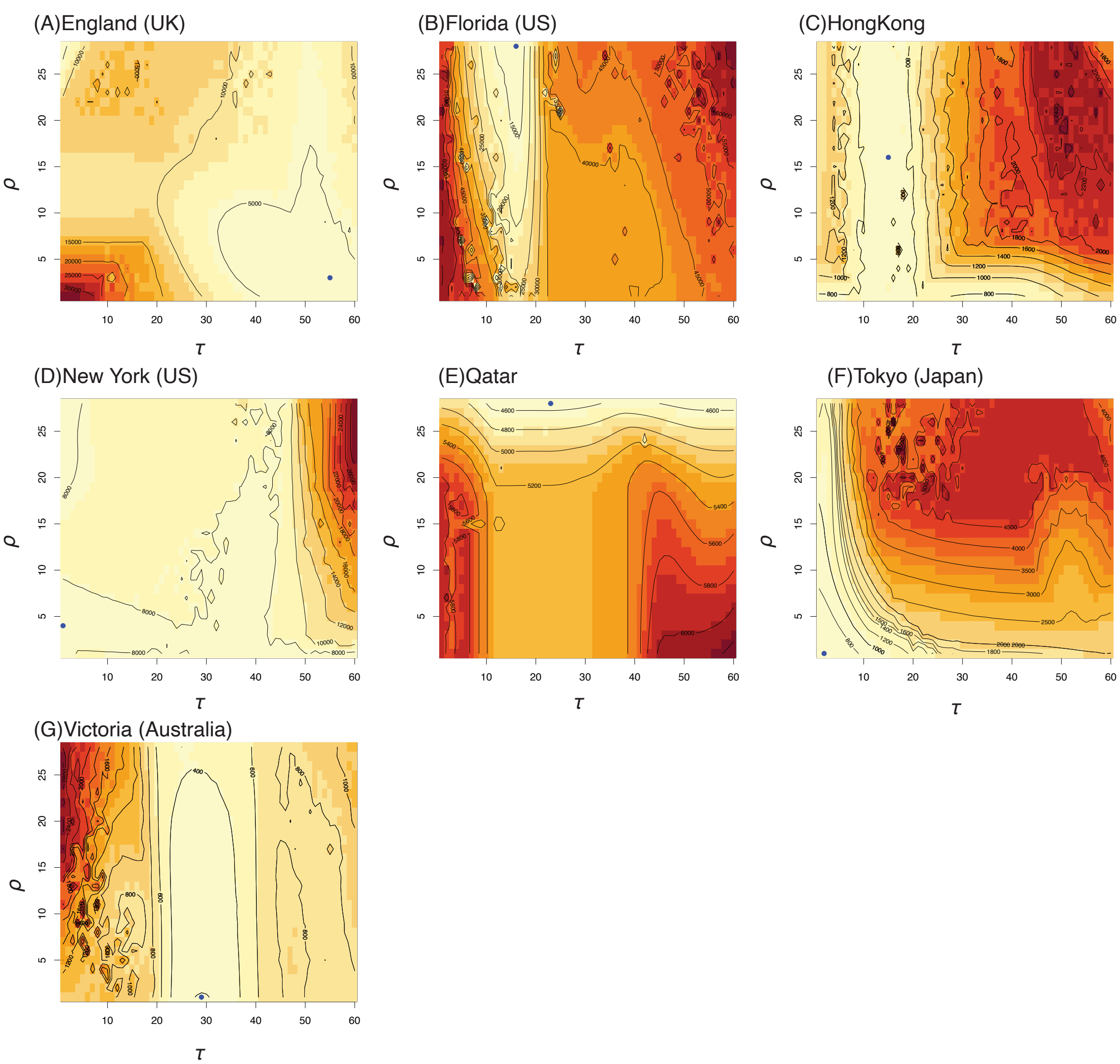

**Supplementary Fig. 1.** Likelihood profile analysis for the risk-mediated transmission model with respect to $\tau$ and $\rho$. Color show negative log likelihood; white denotes low value and black denotes high value, respectively. These profile likelihood landscapes visualize the practical identifiability of the two delay parameters. The diagonal ridges indicate a compensatory relationship between the cognitive lag and the policy implementation delay in the parameter space.

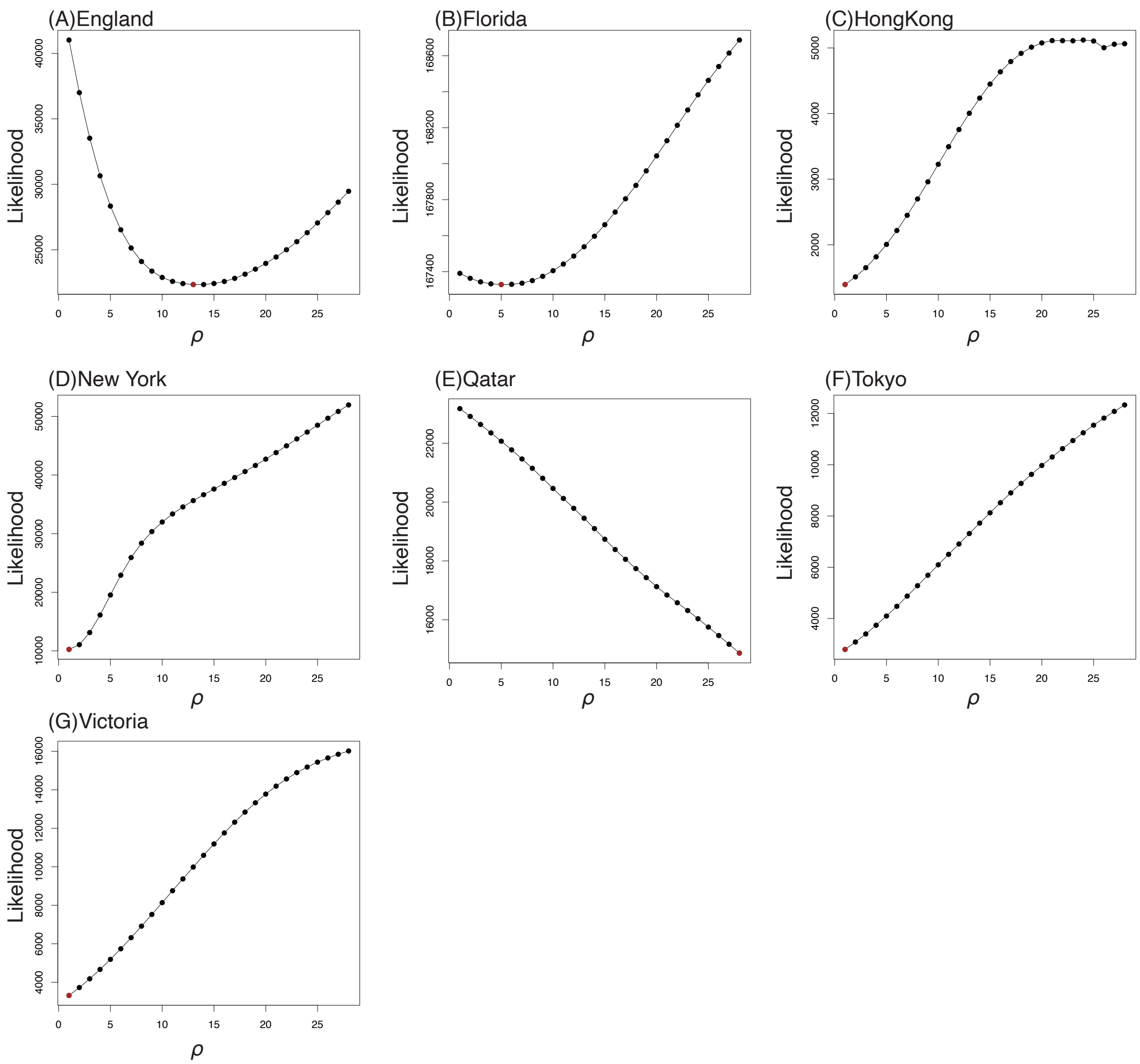

**Supplementary Fig. 2.** Likelihood profile analysis for the baseline model with respect to $\rho$. Note that vertical axis denotes negative log likelihood.

**Supplementary Table**

**Supplementary Table 1.** Parameter estimates and goodness-of-fit metrics for competing models across seven regions. Maximum likelihood estimates of model parameters and their 99% confidence intervals (CI LB denotes lower bound and CI UB denotes upper bound), alongside Akaike Information Criterion (AIC) and Bayesian Information Criterion (BIC) values. RMT, Base, and MB denote the risk-mediated transmission (proposed), baseline, and mobility-based models, respectively. Certain maximum likelihood point estimates fall outside the bootstrap confidence intervals due to compensatory ridges in the parameter space.

| Region | Model | AIC | BIC | $\tau$ | $\tau$ (CI LB) | $\tau$ (CI UB) | $\rho$ | $\rho$ (CI LB) | $\rho$ (CI UB) | $\beta_0$ | $\beta_0$ (CI LB) | $\beta_0$ (CI UB) | $p$ | $p$ (CI LB) | $p$ (CI UB) | $n$ | $n$ (CI LB) | $n$ (CI UB) |
|---|---|---|---|---|---|---|---|---|---|---|---|---|---|---|---|---|---|---|
| England | RMT | 6186.6 | 6302.5 | 55 | 54 | 57 | 3 | 3 | 5 | 0.966 | 0.911 | 0.984 | 0.174 | 0.168 | 0.174 | 49.214 | 44.759 | 49.545 |
| England | Base | 44703.0 | 44761.0 | – | – | – | 13 | 13 | 14 | 0.544 | 0.526 | 0.546 | – | – | – | 1.000 | – | – |
| England | MB | 46071.3 | 46106.1 | – | – | – | – | – | – | 0.524 | 0.524 | 0.525 | – | – | – | 1.000 | – | – |
| Florida | RMT | 22658.1 | 22778.7 | 16 | 14 | 14 | 28 | 28 | 28 | 0.971 | 0.899 | 2.801 | 0.594 | 0.550 | 0.864 | 1.239 | 1.025 | 1.463 |
| Florida | Base | 334666.9 | 334727.2 | – | – | – | 5 | 1 | 1 | 0.439 | 0.439 | 0.440 | – | – | – | 1.000 | – | – |
| Florida | MB | 331968.5 | 332004.7 | – | – | – | – | – | – | 0.461 | 0.460 | 0.462 | – | – | – | 1.000 | – | – |
| Hong Kong | RMT | 1258.0 | 1328.5 | 15 | 1 | 15 | 16 | 1 | 28 | 2.378 | 0.974 | 9.290 | 0.839 | 0.515 | 4.531 | 2.015 | 1.601 | 50.808 |
| Hong Kong | Base | 2798.7 | 2834.0 | – | – | – | 1 | 1 | 1 | 2.720 | 1.445 | 2.915 | – | – | – | 1.000 | – | – |
| Hong Kong | MB | 9950.1 | 9971.3 | – | – | – | – | – | – | 0.447 | 0.422 | 0.451 | – | – | – | 1.000 | – | – |
| New York | RMT | 6207.6 | 12435.1 | 1 | 1 | 1 | 4 | 3 | 3 | 1.162 | 1.161 | 1.426 | 0.090 | 0.088 | 0.752 | 91.290 | 9.109 | 91.290 |
| New York | Base | 10248.8 | 20507.5 | – | – | – | 1 | 1 | 1 | 2.305 | 2.270 | 2.342 | – | – | – | 1.000 | – | – |
| New York | MB | 35727.9 | 71461.8 | – | – | – | – | – | – | 0.511 | 0.510 | 0.511 | – | – | – | 1.000 | – | – |
| Qatar | RMT | 4523.2 | 9066.5 | 23 | 12 | 20 | 28 | 28 | 28 | 0.459 | 0.458 | 1.243 | 0.215 | 0.210 | 0.673 | 87.924 | 2.064 | 94.733 |
| Qatar | Base | 14868.0 | 29746.0 | – | – | – | 28 | 28 | 28 | 0.604 | 0.598 | 0.611 | – | – | – | 1.000 | – | – |
| Qatar | MB | 30616.2 | 61238.3 | – | – | – | – | – | – | 0.393 | 0.392 | 0.393 | – | – | – | 1.000 | – | – |
| Tokyo | RMT | 612.3 | 1244.6 | 2 | 1 | 1 | 1 | 1 | 1 | 4.373 | 2.854 | 5.396 | 0.793 | 0.685 | 0.831 | 1.321 | 1.347 | 1.968 |
| Tokyo | Base | 2795.0 | 5599.9 | – | – | – | 1 | 1 | 1 | 0.772 | 0.764 | 0.780 | – | – | – | 1.000 | – | – |
| Tokyo | MB | 14136.0 | 28277.9 | – | – | – | – | – | – | 0.385 | 0.382 | 0.388 | – | – | – | 1.000 | – | – |
| Victoria | RMT | 187.6 | 395.1 | 29 | 27 | 29 | 1 | 1 | 1 | 0.735 | 0.681 | 0.781 | 0.666 | 0.639 | 0.687 | 1.202 | 1.093 | 1.390 |
| Victoria | Base | 3313.6 | 6637.2 | – | – | – | 1 | 1 | 1 | 6.975 | 6.406 | 7.583 | – | – | – | 1.000 | – | – |
| Victoria | MB | 1464.5 | 2935.0 | – | – | – | – | – | – | 0.865 | 0.850 | 0.881 | – | – | – | 1.000 | – | – |

**Supplementary Table 1. (Continued).**

| Region | Model | $k_a$ | $k_a$ (CI LB) | $k_a$ (CI UB) | $k_b$ | $k_b$ (CI LB) | $k_b$ (CI UB) | $h_a$ | $h_a$ (CI LB) | $h_a$ (CI UB) | $h_b$ | $h_b$ (CI LB) | $h_b$ (CI UB) | $g$ | $g$ (CI LB) | $g$ (CI UB) |
|---|---|---|---|---|---|---|---|---|---|---|---|---|---|---|---|---|
| England | RMT | 3.97E-03 | 3.43E-03 | 4.21E-03 | 2.66E-07 | 6.22E-08 | 3.00E-07 | 59.6 | 59.0 | 62.7 | 1.37E-03 | 1.27E-03 | 1.59E-03 | – | – | – |
| England | Base | – | – | – | – | – | – | 78.4 | 78.2 | 78.8 | 1.03E-02 | 1.01E-02 | 1.23E-02 | – | – | – |
| England | MB | – | – | – | – | – | – | – | – | – | – | – | – | 1.129 | 1.126 | 1.133 |
| Florida | RMT | 9.17E-32 | 0.00 | 2.35E-08 | 3.06E-05 | 6.62E-06 | 4.01E-05 | 4.9 | 0.0 | 7.5 | 8.81E-02 | 5.56E-02 | 8.90E-02 | – | – | – |
| Florida | Base | – | – | – | – | – | – | 2778.5 | 2682.6 | 2878.6 | 4.34E-52 | 2.48E-91 | 7.65E-47 | – | – | – |
| Florida | MB | – | – | – | – | – | – | – | – | – | – | – | – | 0.255 | 0.248 | 0.261 |
| Hong Kong | RMT | 1.21E-04 | 9.47E-09 | 3.17E-02 | 5.01E-08 | 2.10E-13 | 8.96E-06 | 0.0 | 0.0 | 72.4 | 1.02E-01 | 2.27E-138 | 1.33E-01 | – | – | – |
| Hong Kong | Base | – | – | – | – | – | – | 31.7 | 30.4 | 47.7 | 1.44E-05 | 8.69E-21 | 2.76E-05 | – | – | – |
| Hong Kong | MB | – | – | – | – | – | – | – | – | – | – | – | – | 0.484 | 0.051 | 0.567 |
| New York | RMT | 5.79E-18 | 3.56E-19 | 1.08E-03 | 2.88E-04 | 1.92E-35 | 2.89E-04 | 55.7 | 17.1 | 55.8 | 4.45E-04 | 4.44E-04 | 1.52E-01 | – | – | – |
| New York | Base | – | – | – | – | – | – | 37.5 | 37.2 | 37.8 | 1.10E-04 | 1.06E-04 | 1.14E-04 | – | – | – |
| New York | MB | – | – | – | – | – | – | – | – | – | – | – | – | 1.358 | 1.354 | 1.362 |
| Qatar | RMT | 1.59E-02 | 8.18E-113 | 1.71E-02 | 2.09E-04 | 3.91E-06 | 3.95E-04 | 615.1 | 0.0 | 631.6 | 0.00E+00 | 1.95E-144 | 1.07E-01 | – | – | – |
| Qatar | Base | – | – | – | – | – | – | 161.5 | 158.3 | 164.7 | 1.31E-03 | 1.20E-03 | 1.43E-03 | – | – | – |
| Qatar | MB | – | – | – | – | – | – | – | – | – | – | – | – | 0.104 | 0.100 | 0.108 |
| Tokyo | RMT | 6.63E-05 | 1.61E-09 | 1.72E-04 | 1.53E-07 | 1.30E-07 | 4.13E-07 | 35.0 | 34.5 | 36.0 | 3.88E-04 | 3.14E-04 | 4.49E-04 | – | – | – |
| Tokyo | Base | – | – | – | – | – | – | 48.4 | 47.7 | 49.1 | 1.12E-27 | 3.76E-122 | 3.50E-13 | – | – | – |
| Tokyo | MB | – | – | – | – | – | – | – | – | – | – | – | – | 0.045 | 0.012 | 0.078 |
| Victoria | RMT | 1.26E-76 | 6.06E-247 | 4.66E-07 | 8.58E-07 | 7.16E-07 | 1.09E-06 | 2.9E-144 | 5.2E-150 | 7.3E-11 | 3.42E-01 | 2.98E-01 | 4.09E-01 | – | – | – |
| Victoria | Base | – | – | – | – | – | – | 26.9 | 26.2 | 27.6 | 8.75E-64 | 5.65E-143 | 6.67E-13 | – | – | – |
| Victoria | MB | – | – | – | – | – | – | – | – | – | – | – | – | 2.679 | 2.631 | 2.727 |

1